\documentclass[10pt]{article}
\usepackage{fullpage,graphicx,subfigure,mathdots,mathpazo,color}
\usepackage{amsmath,amscd,tikz,mathrsfs,cite}
\usepackage[normalem]{ulem}
\usepackage{amsmath}
\usepackage{setspace}

\usepackage{epsfig,amsmath,graphicx,amssymb,overpic}
\usepackage{bm}
\usepackage{graphicx}
\usepackage{subfigure, xcolor}
\usepackage{ntheorem}
\usepackage{listings,diagbox}

\usepackage{color}
\usepackage{epsfig,amsmath,graphicx,amssymb,overpic}

\usepackage{booktabs}
\usepackage{diagbox}
\usepackage{graphicx}
\usepackage{dcolumn}
\usepackage{bm}
\usepackage{graphicx}
\usepackage{subfigure}
\usepackage{epsfig,amsmath,graphicx,amssymb,overpic,cite}
\usepackage{makecell}

\def\be{\begin{equation}}
\def\ee{\end{equation}}
\def\bee{\begin{eqnarray}}
\def\ene{\end{eqnarray}}
\def\bes{\begin{subequations}}
\def\ees{\end{subequations}}

\newcommand{\PT}{{\cal PT}}

\def\v{\vspace{0.1in}}

\usepackage{geometry}
\allowdisplaybreaks[4]

\begin{document}

\baselineskip=14pt
\renewcommand {\thefootnote}{\dag}
\renewcommand {\thefootnote}{\ddag}
\renewcommand {\thefootnote}{ }

\pagestyle{plain}

\begin{center}
\baselineskip=16pt \leftline{} \vspace{-.3in} {\Large \bf  Spontaneous symmetry breaking and ghost states supported by the fractional nonlinear Schr\"odinger equation with focusing  saturable nonlinearity and $\PT$-symmetric potential} \\[0.2in]
\end{center}

\begin{center}
Ming Zhong$^{a,b}$, Li Wang$^{c,d}$, Pengfei Li$^{e}$, and Zhenya Yan$^{{a,b},{*}}$ \footnote{$^{*}$Corresponding author. {\it Email address}: zyyan@mmrc.iss.ac.cn}  \\[0.15in]
{\it \small$^{a}$KLMM, Academy of Mathematics and Systems Science, Chinese Academy of Sciences, Beijing 100190, China \\
\it \small$^{b}$School of Mathematical Sciences, University of Chinese Academy of Sciences, Beijing 100049, China \\
\it \small$^{c}$Yanqi Lake Beijing Institute of Mathematical Sciences and Applications, Beijing, 101408, China \\
\it \small$^{d}$ Yau Mathematical Sciences Center and Department of Mathematics, Tsinghua University, Beijing, 100084, China\\
\it \small$^{e}$ Department of Physics, Taiyuan Normal University, Taiyuan, 030031, China\\}
\end{center}

\vspace{0.1in}

{\baselineskip=13pt

\noindent {\bf Abstract:}\, We report a novel spontaneous symmetry breaking phenomenon  and ghost states existed in the framework of
the fractional nonlinear Schr\"odinger (FNLS) equation with focusing saturable nonlinearity and $\PT$-symmetric potential. The continuous  asymmetric soliton branch bifurcates from the fundamental symmetric one as the power exceeds some critical value. Intriguingly, the symmetry of fundamental solitons is broken into two branches of asymmetry solitons (alias ghost states) with complex conjugate propagation constants, which is solely in fractional media. 
Besides, the dipole (antisymmetry) and tripole solitons are also studied numerically. Moreover, we analyze the influences of fractional L\'{e}vy index ($\alpha$) and saturable nonlinear parameters (S) on the symmetry breaking of solitons in detail. And the stability of fundamental soliton, asymmetric, dipole and tripole solitons are explored via the linear stability analysis and direct propagations. Moreover, we explore the elastic/semi-elastic collision phenomena between symmetric and asymmetric solitons. Meanwhile,  we find the stable  excitations from the fractional diffraction with saturation nonlinearity to integer-order diffraction with Kerr nonlinearity via the adiabatic excitations of parameters. These results will provide some theoretical basis for the study of spontaneous symmetry breaking phenomena and related physical experiments in the fractional media with $\PT$-symmetric potentials. }

\vspace{12 pt}

 \textbf{As a unique phenomenon existed in the area of various physical scenes, spontaneous symmetry breaking (SSB) of solitons in the nonlinear Schr\"odinger (NLS) equation has attracted much attention in recent years. The symmetry breaking can be found in the $\PT$-symmetric dissipative systems, which was  thought to exist only in conservative systems.  And it has been extensively studied in the NLS equation with $\PT$-symmetric potentials. However, to the best of our knowledge, SSB was not even found in the fractional nonlinear Schr\"odinger (FNLS) equation  with focusing  saturable nonlinearity and $\PT$-symmetric potential before. In this paper, a novel symmetry breaking phenomenon and ghost states are reported in such model. The asymmetric solitons  with complex propagation constants bifurcate from the fundamental symmetric solitons.  The two branches  asymmetric solitons have mutually conjugate propagation constants. And the dipole and tripole localized modes are also found numerically, which can maintain symmetry in the  process of SSB. The dependence  of  symmetry breaking on the fractional L\'{e}vy index $\alpha$ and saturable parameter $S$   are also studied, which indicate the ghost states are solely in the fractional media. Moreover, the stability and dynamics of symmetric and asymmetric solitons are explored via the linear stability analysis, direct simulation, interaction as well as adiabatic excitation. These results will provide some theoretical basis for the study of SSB phenomena and related physical experiments in the fractional media with $\PT$-symmetric potentials. }


\vspace{-0.05in}

\vspace{0.1in}

\section{Introduction}

$\PT$-symmetry has been a topic issue since it was first put forward by Bender and Boettcher~\cite{B98} in 1998. They introduced a family of complex non-Hermitian Hamiltonians ${\mathcal H}=-\partial_x^2-(ix)^{\gamma}$ with $\gamma\geq 2$ which admits the fully real spectra. And a non-Hermitian Hamiltonian $\mathcal{H}$ is said to be $\PT$-symmetric if $[\mathcal{H},\PT]=0$ with $\mathcal{P}$ and $\mathcal{T}$ denoting the parity and time reversal operators, respectively. One can verify that the $\PT$-symmetric condition of a linear Sch\"odinger operator with a complex potential $U(x)$ is that $U(x)=U^{*}(-x)$. Since that, many researches have paid  much attention to the spectra of non-Hermitian $\PT$-symmetric Hamiltonians~\cite{B07,B16,A01,M10,L09,D01}. In general, it is tough to generate continuous families of solitons in dissipative systems~\cite{soli}, however stable solitons can also be found in nonlinear media with $\PT$-symmetric Scarf-II and periodic potentials~\cite{Mu08}. Then plenty of $\PT$-symmetric potentials were introduced into the NLS/NSL-like equations to discover stable solitons~\cite{Ab10,Ab11,Vv12,Ni12,Ac12,Ze12,Bl13,We15,Ko16,Zh16,Ya17,Ya19,Wa19,Zh21,Ch22,So22}, including multi-humps solitons~\cite{Ya17,We15}, peakon solitons~\cite{Ch22,So22,Zh21}, and etc. Moreover, lots of appealing phenomenon have been observed in optical physical experiments~\cite{Ph1,Ph2,Ph3,Ph4,Ph5}. In physical reality, the $\PT$-symmetric potential can be implemented as the complex refractive index with gain and loss distribution in optical experiments.

Spontaneous symmetry breaking (SSB) is a fascinating phenomenon that occurs in various fields of physical systems,  which means the transformations from symmetric or anti-symmetric solitons to asymmetric ones~\cite{SSB1,SSB2}. When the power exceeds the critical value, the system admits symmetric and asymmetric solitons with the same propagation constant. It has been studied extensively in conservative systems with real symmetric potentials~\cite{SSB3,SSB4}. A nature question is that  whether the non-Hermitian optical system with $\PT$-symmetric complex potential would support the asymmetric solitons and SSB phenomenon. In general, since the   bifurcation of solitons needs to satisfy infinitely nontrivial conditions, such symmetry breaking is forbidden~\cite{SSB5}. However, Yang~\cite{SSB6} proposed a special class of one dimensional (1D) $\PT$-symmetric potential $V(x)=g^2(x)+a g(x)+ig'(x)$, where $g(x)$ is  a real even function, $a\in \mathbb{R}$, and the prime denotes the derivative with respet to $x$. With the  special complex potential,   the  SSB phenomenon is supported in the NLS equation. Moreover, the  partial-parity-time ($\mathcal{PPT}$) potential is introduced to NLS equation in two-dimensional (2D) case to find the symmetry breaking~\cite{SSB7}. And then the  SSB phenomena existed in the non-Hermitian systems with $\PT$-symmetric potentials were extended to 1D/2D NLS equations with different types of nonlinearities~\cite{SSB8,SSB9,SSB10,SSB11}.

As a generalization of traditional quantum mechanics, the space-fractional  quantum mechanics (SFQM) was first proposed by Laskin~\cite{FNLS1,FNLS2}. As a result, the underlying fundamental NLS equation was replaced by fractional nonlinear Schr\"odinger (FNLS) equation characterised by the L\'{e}vy index $\alpha\, (1< \alpha\leq 2)$~\cite{FNLS3}
\begin{equation}\label{FNLS-0}
i\frac{\partial\psi}{\partial z}-\left(-\frac{\partial^2}{\partial x^2}\right)^{\!\!\alpha/2}\psi+V(x)\psi
+f(|\psi|^2)\psi=0,
\end{equation}
where $f(|\psi|^2)=\sigma |\psi|^2$,  with $\sigma=\pm 1$, and $V(x)$ is the external potential, the fractional-order derivative implies the existence of unconventional diffraction effects. The FNLS equation describes the evolutionary behavior of fractional spin particles when the Feynman path integral is replaced by L\'{e}vy trajectories~\cite{FNLS1,FNLS2}.
   Despite the current difficulties in the practical application of the FNLS model~\cite{FNLS10,FNLS11}, the relevant experimental schemes have been proposed in condensed matter and optical cavities~\cite{FNLS12,FNLS13}.  Besides, the existence and  dynamical behaviors of solitons in the FNLS equations have become  a research hot-spot~\cite{FNLS4,FNLS5,FNLS6,FNLS7,FNLS8,FNLS9,FNLS14}. And the symmetry breaking can also be found  in the FNLS equations with the real symmetric potentials~\cite{FNLSSSB1,FNLSSSB2,FNLSSSB3}. However, there are few works about the SSB existed in the FNLS equation with the special  $\PT$-symmetric potential~\cite{FNLSSSB4}. Whether the phenomenon of symmetry breaking in the FNLS equation is different from that of integral order NLS equation. This is the main motivation of this paper.

To this end, we investigate the novel symmetry breaking phenomenon existed in the framework of FNLS equation with focusing saturable nonlinearity and $\PT$-symmetric complex potential in detail. The rest of this paper is arranged as follows. In Sec. 2, we introduce the FNLS equation with the focusing saturable nonlinearity and the corresponding $\PT$-symmetric potential briefly. And we investigate the SSB and ghost states in the FNLS equation with focusing saturable nonlinearity and $\PT$ potential in Sec. 3. Moreover, the dipole and tripole solitons for the excited states are also found numerically. We also explore the dependence of symmetry breaking on L\'evy index $\alpha$ and saturable parameter $S$. Moreover, the stability and dynamics of symmetric and asymmetric solitons have been studied from a variety of aspects in Sec. 4. The final conclusions and discussions are made in Sec. 5.

\section{$\PT$-symmetric fractional nonlinear physical model}

In the saturable nonlinear media (e.g.,  semiconductor doped glasses and in photorefractive media~\cite{sn1,sn2}), the propagation of  beams with fractional-order diffraction effect can be described by the following dimensionless FNLS equation~\cite{FNLS14,FNLSSSB3}
\begin{equation}\label{FNLS}
i\frac{\partial\psi}{\partial z}-\left(-\frac{\partial^2}{\partial x^2}\right)^{\!\!\alpha/2}\psi+V(x)\psi+\sigma \frac{|\psi|^2\psi}{1+S|\psi|^2}=0,
\end{equation}
where $\psi=\psi(x,z)$ represents the amplitude of light field, $z$ the propagation distance, $x$ the propagation coordinates, and $\sigma=\pm1$ corresponds to the self-focusing (+) and self-defocusing (-) saturable nonlinearity, respectively. $S>0$ denotes the saturable parameter. As $S=0$, the saturable nonlinear term reduces to the Kerr nonlinear term. Eq.~(\ref{FNLS}) is associated with a variational principle $i\partial\psi/(\partial z)=\delta H/(\delta\psi^*)$ with the Hamiltonian
\bee
H=\int_{\mathbb{R}}\left[\psi^*\left(-\frac{\partial^2}{\partial x^2}\right)^{\!\!\alpha/2}\psi
-[V(x)+\sigma/S]|\psi|^2+\frac{\sigma}{S^2}\ln(1+S|\psi|^2)\right]dx.
\ene

\begin{figure}[!t]
    \centering
\vspace{-0.15in}
  {\scalebox{0.8}[0.8]{\includegraphics{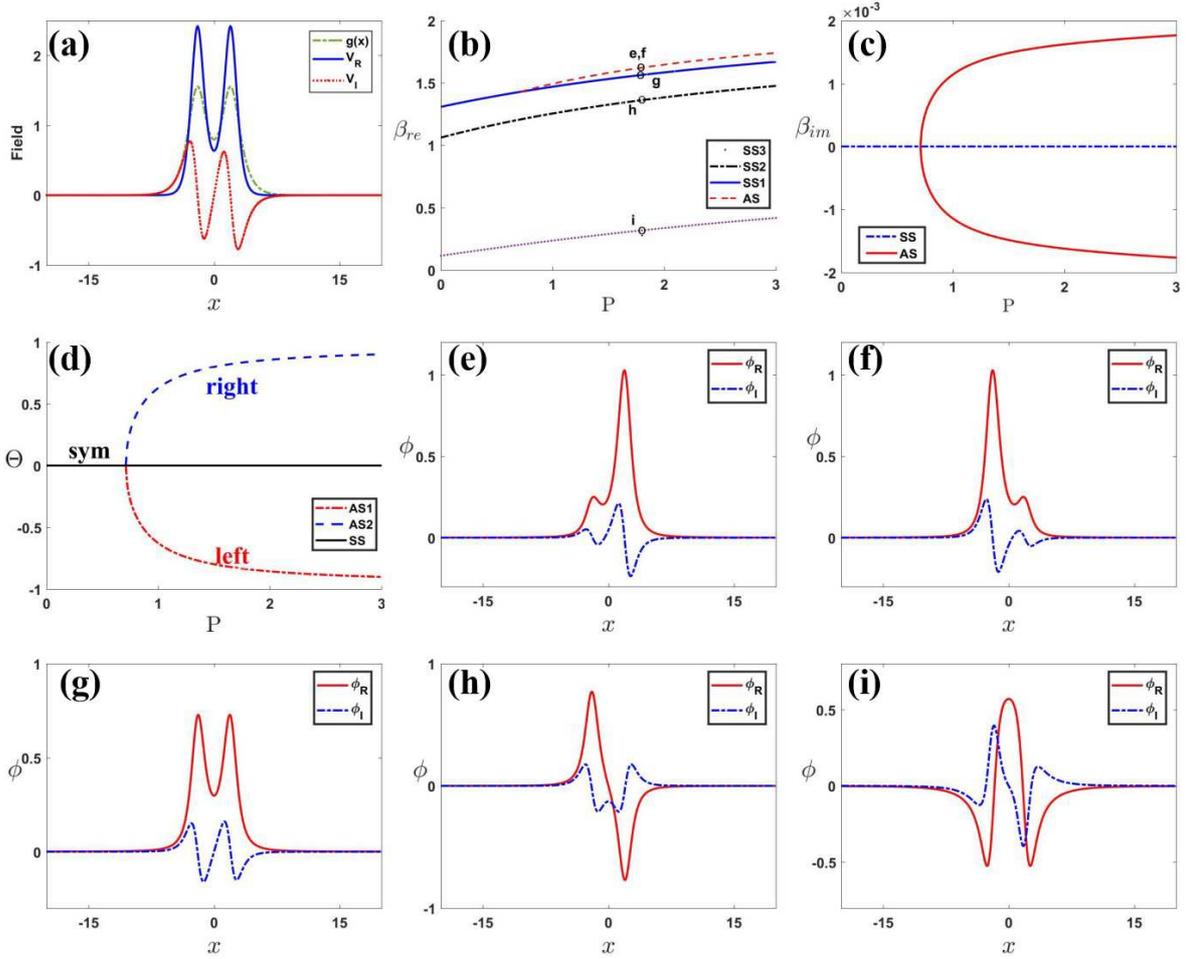}}}\hspace{-0.3in}
\vspace{0.1in}
\caption{\small Symmetry breaking of solitons in Eq.~(\ref{ode}) with L\'evy index $\alpha=1.5$ and saturable parameter $S=1$. (a) The $\PT$-symmetric double-hump potential well given by Eq.~(\ref{PT1}). (b) The real part of the propagation constant $\beta_{re}$ versus power $P$ for the fundamental symmetric, the dipole soliton, the tripole soliton for the excited states as well as the asymmetric soliton (ghost states). (c) The imaginary part of the propagation constant $\beta_{im}$ versus power $P$ for the all symmetric and asymmetric solitons. (d) The dependence of $\Theta$ on the power for the   all symmetric and asymmetric solitons. (e-i) Profiles of symmetric and asymmetric solitons with power $P=1.8$ labeled in Fig.~\ref{SSB}(b). }
  \label{SSB}
\end{figure}

 The focusing case $\sigma=1$ is considered in the following discussions. Here   $(-\frac{\partial^2}{\partial x^2})^{\alpha/2}$ is the fractional Laplacian with $1<\alpha\leq2$ denoting the L\'{e}vy index. One can easily verify that Eq.~(\ref{FNLS}) reduces to the NLS equation when $\alpha=2$. And $V(x)$ represents the complex $\PT$-symmetric potential taken the form as~\cite{SSB6,SSB9,Ya17}
\begin{equation}\label{PT1}
V(x)=g^2(x)+ig'(x),
\end{equation}
with $g(x)$ being an even and real-valued function, and the prime denoting the derivative with respect to $x$. Here we consider $g(x)$ to be a double-hump  potential well
\begin{equation}\label{PT2}
g(x)=V_0\left[{\rm sech}\left(\frac{x-x_0}{w_0}\right)+{\rm sech}\left(\frac{x+x_0}{w_0}\right)\right],
\end{equation}
where $V_0$ and $w_0$ denotes the modulation depth and  width of the well, respectively, and $x_0$ represents the distance between the two peaks of the potential. And one can deduce the real and
imaginary parts of complex potential $V(x)$ via Eq.~(\ref{PT1}). Without loss of generality, we choose $V_0=1.5, x_0=2$ and $w_0=1$. The profile of $g(x)$, the real and imaginary parts of $V(x)$ have depicted in Fig.~\ref{SSB}(a). Notice that one can also choose $g(x)$ as the other double-hump  potential wells.

The localized stationary solutions of Eq.~(\ref{FNLS}) are sought  as $\psi(x,z)=\phi(x; \beta)e^{i\beta z}$, where $\beta=\beta_{re}+i\beta_{im}$ is the complex constant, which is different compared to  the general case. In fact, the solutions with complex constant have been reported in the other physical systems, and were named the spatially asymmetric ghost states~\cite{GS1,GS2,GS3}. And $\phi(x; \beta)$ satisfies the stationary equation
\begin{equation}\label{ode}
-\left(-\frac{\partial^2}{\partial x^2}\right)^{\!\!\alpha/2}\phi+[g^2(x)+ig'(x)]\phi+\sigma \frac{|\phi|^2\phi}{1+S|\phi|^2}=\beta\phi.
\end{equation}
It's noted that we omitted the factor $e^{-2\beta_{im}z}$ due to the fact that $\beta_{im}$ is very small actually, which can be seen in the following discussion. Since the existence of the fractional derivative operator and complex potential, Eq.~(\ref{ode}) is not solvable exactly in general.  Thus the accelerated imaginary-time evolution method (AITEM) and power-conserved squared-operator method (PCSOM)~\cite{Yang} are adopted to obtain the fundamental symmetric, asymmetric, dipole and tripole solitons.  And it's worthy to mention that the fractional   Laplacian operator $(-\frac{\partial^2}{\partial x^2})^{\alpha/2}$ can be implemented as $F^{-1}\left[|k|^{\alpha}F\left[\cdot\right]\right]$, where $F[\cdot]$ and $F^{-1}[\cdot]$ denotes the Fourier  and inverse Fourier transforms, respectively.

And the soliton solution of Eq.~(\ref{ode}) can be featured by the  power (alias mass or norm) $P\left(\beta\right)$, which is  a conservative quantity  of Eq.~(\ref{FNLS}). To measure the degree of the asymmetric soliton, one can divide $P(\beta)$ into two regions, the left and right ones, that is
\begin{equation}\label{Power}
  P(\beta):=\int_{-\infty}^{+\infty}|\phi(x; \beta)|^2dx
  =P_L(\beta)+P_R(\beta)=\int_{-\infty}^{0}|\phi(x; \beta)|^2dx+\int_{0}^{+\infty}|\phi(x; \beta)|^2dx.
\end{equation}
And the parameter $\Theta(\beta)$ can be defined as
\begin{equation}\label{Theta}
\Theta(\beta)=\frac{P_L(\beta)-P_R(\beta)}{P_L(\beta)+P_R(\beta)}.
\end{equation}
Obviously,  if the soliton is symmetric, then $\Theta(\beta)=0$.

\section{Symmetry breaking and ghost states}

In this section, we mainly research the SSB and  ghost states with complex propagation constants $\beta$ existed in  Eq.~(\ref{ode}).

First, we consider the saturable parameter $S=1$ and L\'evy index $\alpha=1.5$ in Eq.~(\ref{ode}).  The fundamental symmetric soliton (labeled with SS1) with two humps bifurcates from the largest discrete eigenvalue of the linear spectral problem $\beta\approx1.3082.$  And the real part of propagation constant $\beta_{re}$ versus  the branch of SS1 is displayed in Fig.~\ref{SSB}(b). As the soliton power increases, a family of asymmetric solitons (labeled with AS) bifurcates at which point the critical power  $P(\beta)\approx0.74\, (\beta\approx1.431)$ (see Fig.~\ref{SSB}(b)). It is worth mentioning that due to the requirement of $\PT$ symmetry of Eq.~(\ref{ode}), another asymmetric soliton must exist, and they satisfy $\PT$ symmetry, that is, $\phi_1^*(x)=\phi_2(-x)$. And at the same value of $\beta$,  they have identical powers. Intriguingly, within the framework of FNLS equation, it is found that the propagation constants of asymmetric solitons are complex-valued, which is  different from the  symmetry breaking  existed in the NLS equation~\cite{SSB9}, and has never been reported. Indeed, the asymmetric solitons with complex propagation constants are not true solutions, therefore they are called ghost states here. As illustrated in Fig.~\ref{SSB}(c), the imaginary parts of the propagation constants of the two branches of asymmetric solitons are conjugate, and the value of the imaginary part increases slightly with the increasing power, and in the order of $10^{-3}$. However, for symmetric solitons, the propagation constant always remains real (see Fig.~\ref{SSB}(c)). Moreover, the symmetric dipole and triple solitons for the excited states  (labeled with SS2, SS3 respectively) are also found numerically (see Fig.~\ref{SSB}(b)). The system admits the dipole solitons for the excited states bifurcated from the second discrete linear spectrum $\beta\approx1.0615$. In contrast to the ground state soliton, the symmetry of the dipole soliton does not change at this point and always maintains its symmetric shape. The branch of symmetric tripole solitons bifurcates from the third linear eigenvalue $\beta\approx0.1142.$
And the symmetry of the tripole soliton is always maintained as the power increases. To measure the dependence of soliton symmetry on power,  we exhibit the change of $\Theta$ in Fig.~\ref{SSB}(d). It is a natural result that we find $\Theta=0$ for all symmetric solitons (including SS1, SS2 as well as SS3). And with the increase of power, for the two asymmetric solitons, they present a tendency to increase in $\Theta$ while maintaining the left-right symmetry (labeled with right and left).

As an example, the profile of fundamental symmetric soliton, the asymmetric soliton, the dipole as  well as the tripole solitons at power $P=1.8$ are presented in Figs.~(\ref{SSB})(e-i), respectively. It can be seen that the two ghost states obey the $\PT$-symmetry while they are not self-symmetric in Figs.~\ref{SSB}(e,f). Most of the power is concentrated on the left or right side. It is worth noting that due to the influence of other waveguide channels, the peaks are not located in the center of $g(x) (x=\pm x_0)$ exactly. Similarly,  the corresponding symmetric solitons are displayed in Figs.~\ref{SSB}(g-i), respectively. Although the ground state soliton and the first excited state soliton are both double-peaked, the symmetries of their real and imaginary parts are just opposite. And the symmetry of the second excited state (tripole soliton) is consistent with that of the ground state soliton, i.e., the real part is an even function and the imaginary part is an odd one.
 \begin{figure}[!t]
    \centering
\vspace{-0.15in}
  {\scalebox{0.5}[0.5]{\includegraphics{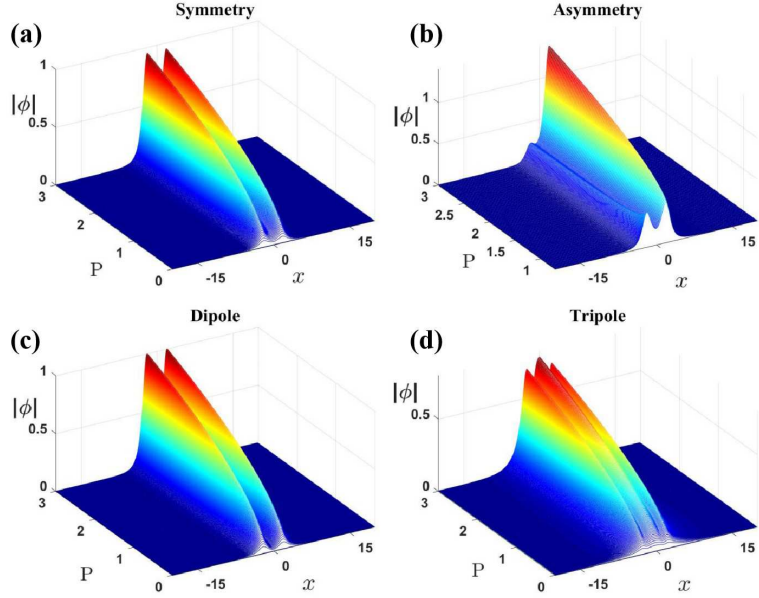}}}\hspace{-0.1in}
\caption{\small Profiles of the  fundamental solitons at (a), asymmetric soliton at (b), dipole soliton and tripole soliton for the excites states at (c) and (d) respectively. The other parameters are the same as Fig.~\ref{SSB}. }
  \label{Intensity}
\end{figure}
The families of symmetric and asymmetric solitons are presented in Fig.~\ref{Intensity}. Fig.~\ref{Intensity}(a) displays the fundamental symmetric double-humps solitons with the power varying from 0.1 to 3. And the symmetric dipole and  tripole solitons are exhibited in Figs.~\ref{Intensity}(c,d), respectively. It is a natural result that the altitudes of the humps increase with the power. As for the branch of  asymmetric solitons bifurcating from the fundamental solution, the altitude of the higher peak is increasing while  the other peak remains almost unchanged with the increasing power.

\begin{figure}[!t]
    \centering
\vspace{-0.15in}
  {\scalebox{0.7}[0.7]{\includegraphics{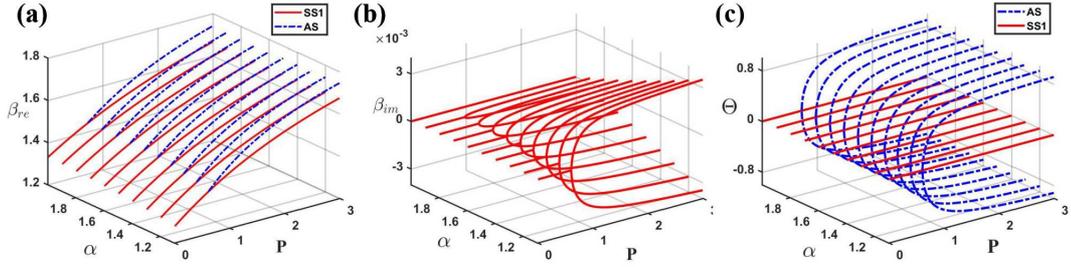}}}\hspace{-0.3in}
\vspace{0.05in}
\caption{\small The dependence of symmetry breaking on L\'evy index $\alpha$. (a) The real part of propagation constant $\beta$. (b) The imagine part of propagation constant $\beta$. (c) The degree of asymmetry $\Theta$. The other parameters are the same as Fig.~\ref{SSB}. }
  \label{alpha}
\end{figure}

We then investigate the dependence of the symmetry breaking bifurcation on  L\'evy index $\alpha$. The real and imaginary parts of the propagation constant versus the power for fundamental symmetric and asymmetric solitons with $\alpha$ varying from 1.1 to 2 are displayed in Fig.~\ref{alpha}(a,b), respectively. It can be seen that the critical bifurcation power changes slightly with the increase  of L\'evy index $\alpha$. Moreover, the real part of the propagation constant is nearly identical for the same power.   However, it can be seen that the imaginary part of $\beta$ decreases with the increase  of $\alpha$ from Fig.~\ref{alpha}(b), which is mainly
due to the fact that the fractional L\'evy index $\alpha$ is significant  for the generation of ghost states. And when $\alpha=2$, the ghost state disappear, i.e., the propagation constants for the branch of asymmetric solitons are real, which illustrates  that the ghost state is a unique phenomenon in the fractional diffraction. We also explore the influence of L\'evy index $\alpha$ on the asymmetry magnitude $\Theta$. It is found that $\Theta$ remains nearly unchanged with the increasing of L\'evy index $\alpha$ (see Fig.~\ref{alpha}(c)).

\begin{figure}[!t]
    \centering
\vspace{-0.15in}
  {\scalebox{0.7}[0.7]{\includegraphics{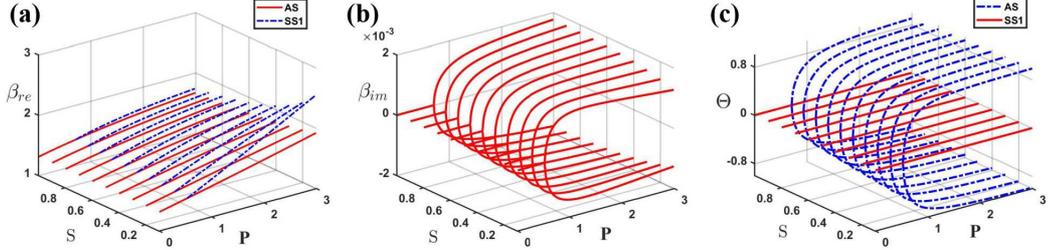}}}\hspace{-0.3in}
\vspace{0.05in}
\caption{\small The dependence of symmetry breaking on saturable parameter $S$. (a) The real part of propagation constant $\beta$. (b) The imagine part of propagation constant $\beta$. (c) The degree of asymmetry $\Theta$. The other parameters are the same as Fig.~\ref{SSB}.}
  \label{S}
\end{figure}

Moreover, we also explore how the stature parameter $S$ effects the SSB bifurcations, and the corresponding results are depicted in Fig.~\ref{S}. It can be seen that the change on real part of propagation constant $\beta$ is greater than that of L\'evy index $\alpha$ from Fig.~\ref{S}(a).  This is due to the change of nonlinearity term in Eq.~(\ref{ode}) with the change of $S$. As $S$ increases, the degree of nonlinearity decreases. However, the imaginary part of $\beta$  changes slightly, this is because of the fact that the fromation of the imaginary part is mainly due to the fractional  diffraction, so $S$ would not have a large impact (see Fig.~\ref{S}(b)). Similar to the previous case, $\Theta$ remains essentially unchanged as displayed in Fig.~\ref{S}(c).

\section{Stability and dynamical behaviors}

In this section, the stability and dynamics of symmetric and asymmetric solitons are explored via the linear stability analysis, direct simulation, interaction as well as adiabatic excitation.

{\it Case 1.---Stability}. We investigate the spectral stability of the obtained numerical solitons by considering the perturbed solutions
\begin{equation}\label{perturbed}
  \psi(x,z)=\left[\phi(x; \beta)+u(x)e^{\delta z}+v^*(x)e^{\delta^*z}\right]e^{i\beta z},
\end{equation}
where $\phi(x; \beta)$ denotes the stationary soliton with the propagation constant $\beta$,  $u(x)$ and $v(x)$ represent  small perturbations with $|u|,|v|\ll|\phi|$, and $\delta$ indicates the growth rate of  instability. By inserting Eq.~(\ref{perturbed}) into Eq.~(\ref{FNLS}) and linearizing with respect to $\delta$, one can obtain   following eigenvalue problem
\begin{equation}\label{spectral}
  i\left(
 \begin{matrix}
   L_1 & L_2  \v\\   -L_2^* & -L_1^*  \end{matrix}  \right)
   \!\!\left( \begin{matrix}   u   \v\\   v  \end{matrix}  \right)
   \!=\!\delta\left( \begin{matrix}   u   \v\\   v  \end{matrix}  \right),
\end{equation}
where
\begin{equation}\label{L1}
L_{1}=-\left(-\frac{\partial^2}{\partial x^2}\right)^{\alpha/2}+g^2(x)+ig'(x)-\beta+\frac{2 \sigma|\phi|^2}{1+S|\phi|^2}-\frac{\sigma S|\phi|^4}{\left(1+S|\phi|^2\right)^2},
\end{equation}
\begin{equation}\label{L2}
L_{2}=\frac{ \sigma \phi^2}{1+S|\phi|^2}-\frac{\sigma S|\phi|^2 \phi^2}{\left(1+S|\phi|^2\right)^2}.
\end{equation}
The above spectral problem can be solved by  means of Fourier collocation method~\cite{Yang}. One can easily observe that the solution $\phi(x)$  is stable if and  only if Re$(\delta)=0$.

\begin{figure}[!t]
    \centering
\vspace{-0.15in}
  {\scalebox{0.6}[0.6]{\includegraphics{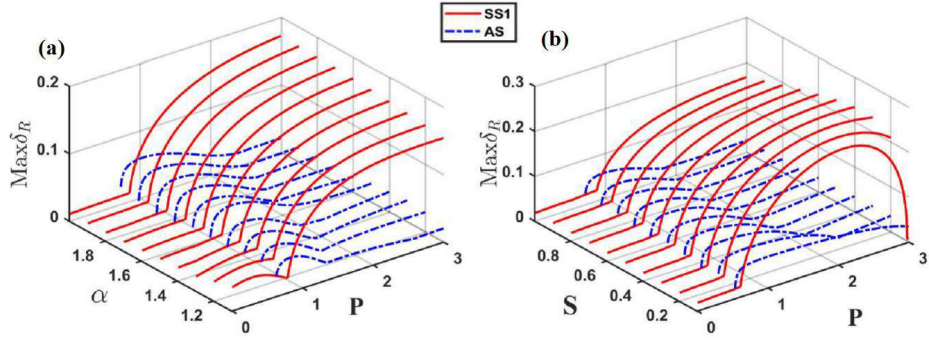}}}\hspace{-0.3in}
\vspace{0.05in}
\caption{\small The dependence of maximal real part of linear spectral of Eq.~(\ref{spectral}) on L\'evy index $\alpha$ in (a) and saturable parameter $S$ in (b). The other parameters are the same as Fig.~\ref{SSB}. }
  \label{MaxImag}
\end{figure}

First, we investigate the dependence of maximal real part of $\delta$ on the L\'evy index $\alpha$ for the fundamental symmetric and asymmetric solitons. The corresponding numerical results are depicted in Fig.~\ref{MaxImag}(a). It's found that the symmetric soliton is weakly linear stable below the critical power $P$ for $\alpha$ ranging from 1.1 to 2. And after the bifurcation of asymmetric solitons, the fundamental solitons is  weakly linear stable only for a narrow range of power, and then become linear unstable immediately. The ghost states is  weakly linear stable for a narrow range of power, which can be seen from Fig.~\ref{MaxImag}(a). It is the fact that due to the complexity of model Eq.~(\ref{FNLS}),  the linear stability is more complicated compared to the other models~\cite{SSB7,SSB8}. Besides, the effect of saturable parameter $S$ on the linear stability analysis is also considered in Fig.~\ref{MaxImag}(b). It can be concluded that the case is similar with the change of $\alpha$. The fundamental symmetric solitons become linear unstable after the existence of ghost states, and the asymmetric branch solitons are mostly unstable.

\begin{figure}[!t]
    \centering
\vspace{-0.15in}
  {\scalebox{0.8}[0.8]{\includegraphics{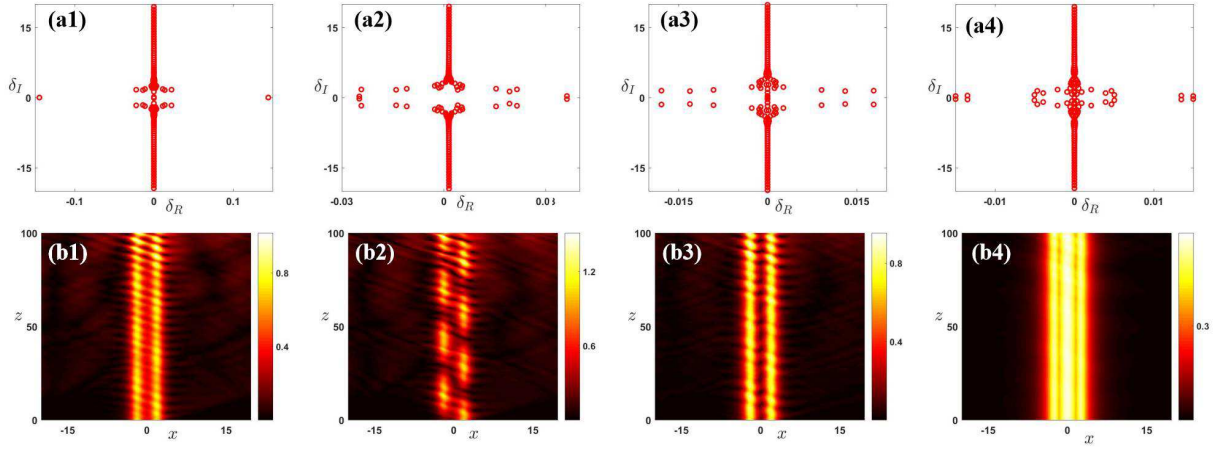}}}\hspace{-0.3in}
\vspace{0.05in}
\caption{\small Linear spectra (the first row) and corresponding evolutions (the second row). (a1,b1) The fundamental symmetric soliton in Fig.~\ref{SSB}(g). (a2,b2) The asymmetric soliton  in Fig.~\ref{SSB}(e). (a3,b3) The dipole soliton in Fig.~\ref{SSB}(h). (a4,b4) The tripole soliton in Fig.~\ref{SSB}(i). Other parameters are the same as Fig.~\ref{SSB}. }
  \label{Dynamic}
\end{figure}

These results are also confirmed via employing the direct simulations of the numerical solitons in Figs.~\ref{SSB}(e-i) $(P=1.8)$.  And we display the linear stability result in the left column in Fig.~\ref{Dynamic} while the right column exhibits the dynamic behaviour. For the fundamental symmetric solitons with $P=1.8$, the linear stability result indicates the soliton is linear unstable (see Fig.~\ref{Dynamic}(a1)), and the propagation with 5\% added to the initial condition also displays the unstable evolution (see Fig.~\ref{Dynamic}(b1)). It can be seen that the oscillation between the two humps and then diverges.   As for the ghost states, we also investigate the corresponding linear spectral and dynamic propagation with $5\%$ random noise. The linear eigenvalue denotes the unstable behaviour in Fig.~\ref{Dynamic}(a2) and then confirmed by the direct simulation (see Fig.~\ref{Dynamic}(b2)). The other asymmetric soliton performs the same unstable behavior, which is not exhibited here. And the weakly unstable linear spectra and dynamic behaviors of the dipole and tripole solitons for the excitation states are also presented in Figs.~\ref{Dynamic}(a3-b4).

\begin{figure}[!t]
    \centering
\vspace{-0.15in}
  {\scalebox{0.8}[0.8]{\includegraphics{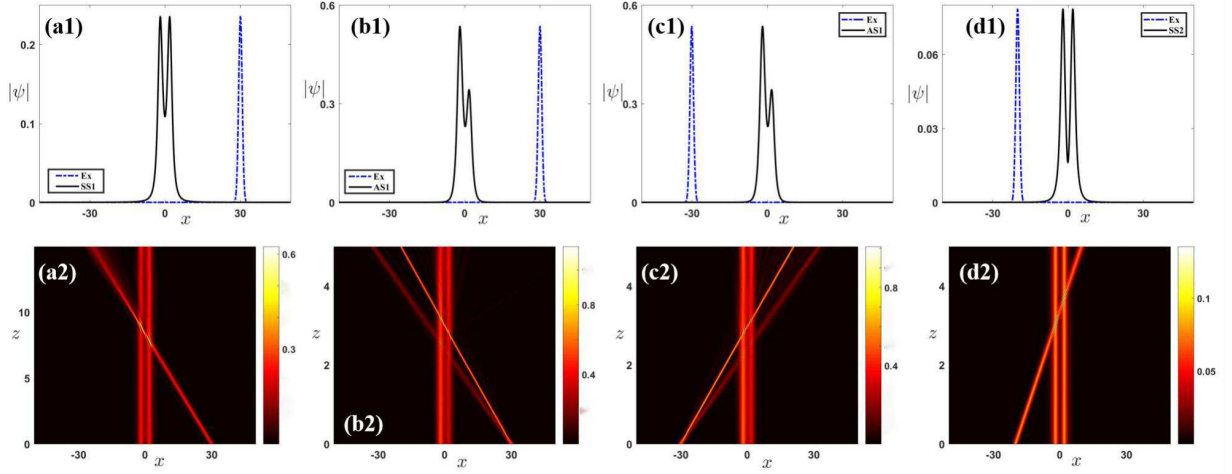}}}\hspace{-0.3in}
\vspace{0.1in}
\caption{\small (a1-a2) The interaction between fundamental soliton with $P=0.2,\alpha=1.5$ and outside solitary wave.  (b1-b2) The interaction between asymmetric soliton $P=0.8,\alpha=1.9$ and outside solitary wave located in the right.  (c1-c2) The interaction between asymmetric  soliton with $P=0.8,\alpha=1.9$ and outside solitary wave located in the left. (d1-d2) The interaction between dipole soliton with $P=0.02,\alpha=1.5$ and outside solitary wave. The other parameters are the same as Fig.~\ref{SSB}. }
  \label{interaction}
\end{figure}

{\it Case 2.---Soliton interactions}. We also  examine the interactions of the symmetric and asymmetric solitons with the exotic isolated solitary wave. For $P=0.2$ and $\alpha=1.5$, we consider the interaction of  fundamental symmetric solitons with the outside Gaussian solitary wave, i.e. the initial condition is chosen as (see Fig.~\ref{interaction}(a1))
\begin{equation}\label{In1}
  \psi(x,0)=\phi(x)+ 0.2354e^{-(x-30)^2-6ix},
\end{equation}
where the value $0.2354$ represents the maximal amplitude of $\phi(x)$ (similar hereinafter). It can be seen that the elastic interaction between them from Fig.~\ref{interaction}(a2), i.e., the shape and intensity of the double-hump solution and the exotic isolated wave remain unchanged before and after the collision. We then choose $P=0.8$ and $\alpha=1.9$ to find the collision phenomenon between the asymmetric soliton with the outside wave, and the initial condition is chosen as (see Fig.~\ref{interaction}(b1))
\begin{equation}\label{In2}
  \psi(x,0)=\phi(x)+ 0.5354e^{-(x-30)^2-8ix}.
\end{equation}
In contrast to the previous case, it can be found  that the Gaussian wave is immediately split into two waves with different phases, and propagates forward, then successively interacts with the asymmetric soliton in Fig.~\ref{interaction}(b2). And these three waves keep  their shapes before and after the interaction. To further explore this phenomenon, we place the exotic Gaussian wave to the left of the asymmetric soliton, i.e., the initial condition is chosen as (see Fig.~\ref{interaction}(c1))
\begin{equation}\label{In3}
  \psi(x,0)=\phi(x)+ 0.5354e^{-(x+30)^2+8ix}.
\end{equation}
And the same interaction is displayed in Fig.~\ref{interaction}(c2). This may be due to the effect of fractional diffraction. Lastly, we consider the dipole soliton for the excites state with $P=0.02$ and $\alpha=1.5$. The initial condition is chosen as (see Fig.~\ref{interaction}(d1))
\begin{equation}\label{In4}
  \psi(x,0)=\phi(x)+ 0.0784e^{-(x+20)^2+80ix}.
\end{equation}
And it can also observe that the elastic collision between them in Fig.~\ref{interaction}(d2).

\v {\it Case 3.---Soliton adiabatic excitations}. We then change the two parameters $\alpha$ and $S$ in the FNLS Eq.~(\ref{FNLS}) as the functions of propagation  distance $z$, that is, we apply an adiabatic switch~\cite{Ya15, We15} to the FNLS equation:
$\alpha\rightarrow\alpha(z)$ and $S\rightarrow S(z)$ with $\alpha(z),S(z)$ being taken as
\begin{equation}\label{condition}
\varrho(z)\!=\!\left\{\!
\begin{aligned}
& \frac{\varrho_2-\varrho_1}{2}\left[1-\cos\left(\frac{2\pi z}{z_{\text{max}}}\right)\right]+\varrho_1,           & 0\leq z < \frac{z_{\text{max}}}{2},\\
&\varrho_2, & z\geq\frac{z_{\text{max}}}{2},
\end{aligned}
\right.
\end{equation}
with $\varrho_{1,2}$ representing the initial and final values in the excitation, respectively, and $z_{\text{max}}$ the maximal value of numerical propagation. And at this point, Eq.~(\ref{FNLS}) is also changed into
\begin{equation}\label{FNLS2}
i\frac{\partial\psi}{\partial z}-\left(-\frac{\partial^2}{\partial x^2}\right)^{\!\!\alpha(z)/2}\psi+[g^2(x)+ig'(x)]\psi+\sigma \frac{|\psi|^2\psi}{1+S(z)|\psi|^2}=0,
\end{equation}
with $\alpha(z)$ and $S(z)$ defined as Eq.~(\ref{condition}).
\begin{figure}[!t]
    \centering
\vspace{-0.15in}
  {\scalebox{0.65}[0.65]{\includegraphics{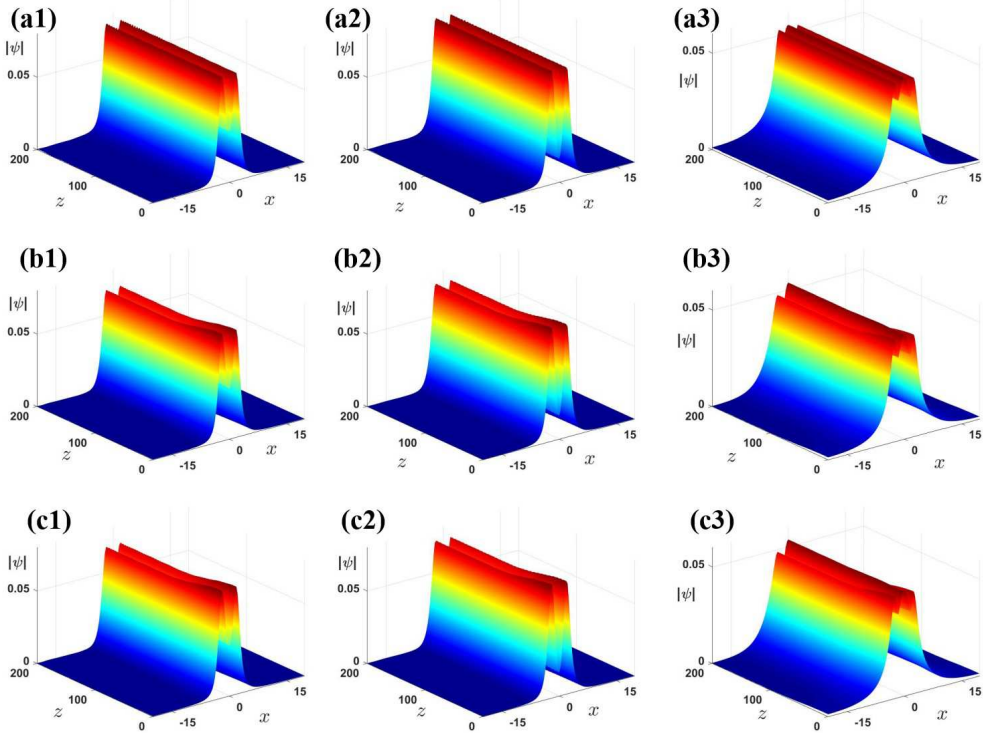}}}\hspace{-0.1in}
\vspace{0.05in}
\caption{\small The first column displays the excitation of fundamental symmetric soliton, the second column displays that of dipole soliton and the third column displays that of tripole soliton. (a1-a3) $\alpha(z)\equiv1.5,S_1=1,S_2=0$.  (b1-b3) $\alpha_1=1.5,\alpha_2=2,S(z)\equiv1$. (a1-a3) $\alpha_1=1.5,\alpha_2=2,S_1=1,S_2=0$. The other parameters are the same as Fig.~\ref{SSB}. }
  \label{excitation}
\end{figure}

{\it Case 3.1}---We consider the adiabatic excitation for the single parameter $\alpha$ or $S$. For the switch of the saturable parameter $S$, the symmetric fundamental soliton, dipole soliton as well as the tripole soliton with $P=0.02$ (similar hereinafter) and $\alpha(z)\equiv1.5$ are chosen to propagate under the change of $S$. And we choose $S_1=1,\, S_2=0,$ that is, we change the nonlinearity from the saturable case to the Kerr one. It is found that the amplitudes of all solitons keep unchanged in these processes from Figs.~\ref{excitation}(a1-a3). We then fix $S(z)\equiv1$, and choose $\alpha_1=1.5,\, \alpha_2=2$, that is, the diffraction are changed from the fractional case to the integer-order one. It can be seen that the humps of fundamental soliton and dipole soliton both decrease at first, and then keep unchanged in the last propagation processes (see Figs.~\ref{excitation}(b1,b2)). However, we find the middle hump of the tripole soliton increases at first, then decreases and keep remain at last (see Fig.~\ref{excitation}(b3)).

{\it Case 3.2}---We now apply the adiabatic switch to both parameters simultaneously, that is, the parameters are chosen as $\alpha_1=1.5,\, \alpha_2=2$ and $S_1=1,\, S_2=0.$ We change the model from the fractional diffraction with saturation nonlinearity to the integer-order diffraction with Kerr nonlinearity, which can be considered as a simple superposition of the first two single-parameter excitations.  And the process exhibited in Figs.~\ref{excitation}(c1-c3) for symmetric soltions are similar with the previous cases. These results may be useful the designs of relevant experiments in fractional media.

\section{Conclusions and discussions}

In summary, we have found a kind of novel symmetry breaking phenomena and ghost states existed in the framework of FNLS equation with focusing saturable nonlinearity and $\PT$-symmetric potential.  The branches of fundamental solitons, asymmetric solitons, the dipole as well as the tripole solitons for the excited state are found numerically. And the asymmetric solitons with complex propagation constant bifurcates  from the branch of fundamental soliton, while the excited states keep symmetry. And we investigate the influence of L\'evy index $\alpha$ and saturable parameter $S$ on the symmetry breaking and ghost state. It is found that L\'evy index $\alpha$ influences the imaginary part of propagation constant and saturable parameter $S$ mainly have effects on power $P$. Moreover, we explore the stability of numerical solitons via linear stability analysis, direct simulation, interaction and excitation. These results may provide the theoretical support for the optical experiments.

Moreover, these results can also be naturally extended to the fractional NLS with the cubic-quintic competing saturable
nonlinearities and $\PT$-symmetric potential~\cite{SSB11}
\begin{equation}\label{FNLS-3-5}
i\frac{\partial\psi}{\partial z}-\left(-\frac{\partial^2}{\partial x^2}\right)^{\!\!\alpha/2}\psi+V(x)\psi+\sigma_1 \frac{|\psi|^2\psi}{1+S|\psi|^2}++\sigma_2 \frac{|\psi|^4\psi}{1+S|\psi|^4}=0,
\end{equation}
where $V(x)$ is the $\PT$-symmetric potential, $\sigma_{1,2}=\pm 1$ correspond to self-focusing or defocusing cubic
and quintic saturable nonlinearities, respectively. Furthermore, whether the ghost state can support by the high-dimensional FNLS equation
\begin{equation}\label{FNLS-h}
i\frac{\partial\psi}{\partial z}-\left(-\nabla^2\right)^{\!\!\alpha/2}\psi+V(x)\psi
+f(|\psi|^2)\psi=0,\quad \nabla^2=\partial_x^2+\partial_y^2,
\end{equation}
and why it can only exist in the fractional diffraction are still open questions to be solved and discussed.

\v\noindent{\bf Acknowledgment} \v

The work  was supported by the NSFC under Grant No. 11925108.


\begin{thebibliography}{0}\setlength{\itemsep}{-0.8mm}
\small
\bibitem{B98}  C. M. Bender and S. Boettcher, Real spectra in non-Hermitian Hamiltonians having PT symmetry. Phys. Rev. Lett. 80, 5243 (1998).

\bibitem{A01} Z. Ahmed, Real and complex discrete eigenvalues in an exactly solvable one- dimensional complex PT-invariant potential. Phys. Lett. A 282, 343 (2001).

\bibitem{D01} P. Dorey,  C. Dunning and  R. Tateo, Spectral equivalences, Bethe ansatz equations, and reality properties in PT-symmetric quantum mechanics.    J. Phys. A 34, 5679 (2001).

\bibitem{B07} C. M. Bender,   Making sense of non-Hermitian Hamiltonians.    Rep. Prog. Phys. 70, 947 (2007).

\bibitem{B16} C. M. Bender,   Rigorous backbone of PT-symmetric quantum mechanics.    J. Phys. A 49, 401002 (2016).

\bibitem{L09}  S. Longhi, Bloch oscillations in complex crystals with PT symmetry. Phys. Rev. Lett. 103, 123601 (2009).

\bibitem{M10} B. Midya, B. Roy, and R. Roychoudhury, A note on the PT invariant periodic potential $V(x)=4\cos^2x+4iV_0\sin 2x$. Phys. Lett. A 374, 2605 (2010).

\bibitem{Mu08}  Z. Musslimani, K. G. Makris, R. El-Ganainy, and D. N. Christodoulides,   Optical solitons in PT periodic potentials.    Phys. Rev. Lett. 100, 030402 (2008).

\bibitem{soli} N. Akhmediev and A. Ankiewicz (eds), Dissipative Solitons (Springer, Berlin, 2005).

\bibitem{Ab10} F. K. Abdullaev, V. V. Konotop, M. Salerno, and A. V. Yulin, Dissipative periodic waves, solitons, and breathers of the nonlinear Schr\"odinger equation with complex potentials. Phys. Rev. E 82, 056606 (2010).

\bibitem{Ab11} F. K. Abdullaev, Y. V. Kartashov, V. V. Konotop, and D. A. Zezyulin, Solitons in PT-symmetric nonlinear lattices. Phys. Rev. A 83, 041805(R) (2011).

\bibitem{Vv12}  D. A. Zezyulin and V. V. Konotop, Nonlinear modes in finite-dimensional PT-symmetric systems. Phys. Rev. Lett. 108, 213906 (2012).

\bibitem{Ni12} S. Nixon, Y. Zhu, and J. Yang, Nonlinear dynamics of wave packets in parity-time-symmetric optical lattices near the phase transition point. Opt. Lett. 37, 4874 (2012).

\bibitem{Ac12}V. Achilleos, P. Kevrekidis, D. Frantzeskakis, and R. Carretero-Gonzalez,   Dark solitons and vortices in PT-symmetric nonlinear media: From spontaneous symmetry breaking to nonlinear PT phase transitions.   Phys. Rev. A 86, 013808 (2012).

\bibitem{Ze12} D. A. Zezyulin and V. V. Konotop,   Nonlinear modes in the harmonic PT-symmetric potential.   Phys. Rev. A 85, 043840 (2012).


\bibitem{Bl13} Y. V. Bludov, V. V. Konotop, and B. A. Malomed, Stable dark solitons in PT-symmetric dual-core waveguides. Phys. Rev. A 87, 013816 (2013)


\bibitem{We15}  Z. Yan, Z. Wen, and V. V. Konotop,   Solitons in a nonlinear Schr\"{o}dinger equation with PT-symmetric potentials and inhomogeneous nonlinearity: Stability and excitation of nonlinear modes.    Phys. Rev. A 92, 023821 (2015).

\bibitem{Ko16} V. V Konotop, J. Yang, and D. A Zezyulin, Nonlinear waves in PT-symmetric systems. Rev. Mod. Phys. 88, 035002 (2016).

\bibitem{Zh16} K. Zhan, H. Tian, X. Li, X. Xu, Z. Jiao, and Y. Jia,  Solitons in PT-symmetric periodic systems with the logarithmically saturable nonlinearity. Sci. Rep. 6, 1 (2016).


\bibitem{Ya17} Z. Yan and Y. Chen, The nonlinear Schr\"odinger equation with generalized nonlinearities and PT-symmetric potentials: Stable solitons, interactions, and excitations. Chaos 27, 073114 (2017).




\bibitem{Ya19} Y. Chen, Z. Yan, and D. Mihalache,  Stable flat-top solitons and peakons in the PT-symmetric $\delta$-signum potentials and nonlinear media. Chaos 29, 083108 (2019).

\bibitem{Wa19} L. Wang, B. A. Malomed, and Z. Yan,  Attraction centers and parity-time-symmetric delta-functional dipoles in critical and supercritical self-focusing media. Phys. Rev. E 99, 052206 (2019).

\bibitem{Ch22} Y. Chen, J. Song, X. Li, and Z. Yan,  Stability and modulation of optical peakons in self-focusing/defocusing Kerr nonlinear media with PT-$\delta$-hyperbolic-function potentials. Chaos 32, 023122 (2022).

\bibitem{Zh21} M. Zhong, Y. Chen, Z. Yan, and S. F. Tian.  Formation, stability, and adiabatic excitation of peakons and double-hump solitons in parity-time-symmetric Dirac-$\delta(x)$-Scarf-II optical potentials. Phys. Rev. E. 105, 014204 (2022).

\bibitem{So22} J. Song, Z. Zhou, W. Weng, and Z. Yan, PT-symmetric peakon solutions in self-focusing/defocusing power-law nonlinear media: Stability, interactions and adiabatic excitations. Physica D 435, 133266 (2022).

\bibitem{Ph1} A. Guo, G. Salamo, D. Duchesne, R. Morandotti, M. Volatier-Ravat, V. Aimez, G. Siviloglou, and D. Christodoulides,   Observation of PT-symmetry breaking in complex optical potentials.    Phys. Rev. Lett. 103, 093902 (2009).

\bibitem{Ph2}  C. E. R\"uter, K. G. Makris, R. El-Ganainy, D. N. Christodoulides, M. Segev, and D. Kip,   Observation of parity-time symmetry in optics.   Nat. Phys. 6, 192 (2010).

\bibitem{Ph3} A. Regensburger, C. Bersch, M.-A. Miri, G. Onishchukov, D. N. Christodoulides, and U. Peschel,   Parity-time synthetic photonic lattices.    Nature 488, 167 (2012).

\bibitem{Ph4} B. Peng, S. K. \"Odemir, F. Lei, F. Monifi, M. Gianfreda, G. L. Long, S. Fan, F. Nori, C. M. Bender, and L. Yang,   Parity-time-symmetric whispering-gallery microcavities. Nature Phys. 10, 394-398 (2014).

\bibitem{Ph5}  H. Hodaei, A. U. Hassan, S. Wittek, H. Garcia-Gracia, R. ElGanainy, D. N. Christodoulides, and M. Khajavikhan, Enhanced sensitivity at higher-order exceptional points. Nature 548, 187 (2017).

\bibitem{SSB1}  B. A. Malomed (ed.), Spontaneous Symmetry Breaking, Self-Trapping, and Josephson Oscillations (Springer, Berlin, 2013).

\bibitem{SSB2}B. A. Malomed and D. Mihalache, Nonlinear waves in optical and matter-wave media: a topical survey of recent theoretical and experimental results. Rom. J. Phys, 64, 106 (2019).

\bibitem{SSB3} A. Sacchetti, Universal critical power for nonlinear Schr\"odinger equations with a symmetric double well potential. Phys. Rev. Lett. 103, 194101 (2009).

\bibitem{SSB4} J. Yang, No stability switching at saddle-node bifurcations of solitary waves in generalized nonlinear Schr\"odinger equations. Phys. Rev. E. 85, 037602 (2012).

\bibitem{SSB5} J. Yang,  Can parity-time-symmetric potentials support families of non-parity-time-symmetric solitons ? Stud. Appl. Math. 132, 332 (2014).

\bibitem{SSB6} J. Yang, Symmetry breaking of solitons in one-dimensional parity-time-symmetric optical potentials. Opt. Lett, 39, 5547 (2014).

\bibitem{SSB7}  J. Yang,  Symmetry breaking of solitons in two-dimensional complex potentials. Phys. Rev. E. 91, 023201 (2015).

\bibitem{SSB8} P. Li and D. Mihalache,   Symmetry breaking of solitons in PT-symmetric potentials with competing cubic-quintic nonlinearity. Proc. Rom. Acad. A 19, 61 (2018).

\bibitem{SSB9} P. Li, C. Dai, R. Li, and Y. Gao,  Symmetric and asymmetric solitons supported by a $\PT$-symmetric potential with saturable nonlinearity: bifurcation, stability and dynamics. Opt. Exp. 26, 6949 (2018).

\bibitem{SSB10} L. Dong, C. Huang, and W. Qi,   Symmetry breaking and restoration of symmetric solitons in partially parity-time-symmetric potentials. Nonlinea. Dyn. 98, 1701 (2019).

\bibitem{SSB11} W. B. Bo, R. R. Wang, W. Liu, and Y. Y. Wang,  Symmetry breaking of solitons in the PT-symmetric nonlinear Schr\"odinger equation with the cubic-quintic competing saturable nonlinearity. Chaos 32, 093104 (2022).

\bibitem{FNLS1} N. Laskin, Fractional quantum mechanics. Phys. Rev. E 62, 3135 (2000).

\bibitem{FNLS2} N. Laskin, Fractional quantum mechanics and L\'{e}vy path integrals. Phys. Lett. A 268 298 (2020).

\bibitem{FNLS3} N. Laskin, Fractional Schr\"odinger equation. Phys. Rev. E 66, 056108 (2002).

\bibitem{FNLS10} Y. C. Wei,  Comment on fractional quantum mechanics and fractional Schr\"{o}dinger equation, Phys. Rev. E 93, 066103 (2016).

\bibitem{FNLS11} N. Laskin, Reply to ``comment on fractional quantum mechanics and fractional Schr\"odinger equation", Phys. Rev. E 93, 066104 (2016).

\bibitem{FNLS12} B. Stickler, Potential condensed-matter realization of space-fractional quantum mechanics: the one-dimensional L\'evy crystal. Phys. Rev. E 88, 12120 (2013).

\bibitem{FNLS13} S. Longhi. Fractional Schr\"odinger equation in optics. Opt Lett. 40, 1117 (2015).

\bibitem{FNLS4} C. Huang and L. Dong, Gap solitons in the nonlinear fractional Schr\"odinger equation with an optical lattice. Opt. Lett. 41, 5636 (2016).

\bibitem{FNLS5} X. Yao and X. Liu, Off-site and on-site vortex solitons in space-fractional photonic lattices. Opt. Lett. 43, 5749 (2018).

\bibitem{FNLS6} X. Yao and X. Liu, Solitons in the fractional Schr\"odinger equation with parity-time-symmetric lattice potential. Photonics Res. 6, 875 (2018).

\bibitem{FNLS7} M. Chen, S. Zeng, D. Lu, W. Hu, and Q. Guo, Optical solitons, self-focusing, and wave collapse in a space-fractional Schr\"{o}dinger equation with a Kerr-type nonlinearity. Phys. Rev. E 98, 022211 (2018).

\bibitem{FNLS8} L. Dong, C. Huang, and W. Qi, Nonlocal solitons in fractional dimensions, Opt. Lett. 44, 4917 (2019).

\bibitem{FNLS9} J. Xie, X. Zhu, and Y. He, Vector solitons in nonlinear fractional Schr\"odinger equations with parity-time-symmetric optical lattices. Nonlinear. Dyn. 97, 1287 (2019).

\bibitem{FNLS14} B. A. Malomed,  Optical solitons and vortices in fractional media: a mini-review of recent results. Photonics, 8, 353(2021).

\bibitem{FNLSSSB1} P. Li and C. Dai,  Double loops and pitchfork symmetry breaking bifurcations of optical solitons in nonlinear fractional Schr\"odinger equation with competing cubic-quintic nonlinearities. Ann. Phys. 532, 2000048 (2020).

\bibitem{FNLSSSB2} P. Li, B. A. Malomed and D. Mihalache,  Symmetry breaking of spatial Kerr solitons in fractional dimension. Chaos, Solitons \& Fractals 132, 109602 (2020).

\bibitem{FNLSSSB3} P. Li, R. Li. and C. Dai, Existence, symmetry breaking bifurcation and stability of two-dimensional optical solitons supported by fractional diffraction. Opt. Exp. 29, 3193 (2021).

\bibitem{FNLSSSB4} P. Li, B. A. Malomed. and D. Mihalache, Symmetry-breaking bifurcations and ghost states in the fractional nonlinear Schr\"{o}dinger equation with a PT-symmetric potential. Opt. Lett. 46, 3267 (2021).


\bibitem{sn1} J. L. Coutaz and M. Kull, Saturation of the nonlinear index of refraction in semiconductor-doped glass.
  J. Opt. Soc. Am. B 8, 95–98 (1991).

\bibitem{sn2} V. Tikhonenko, J. Christou, and B. Luther-Davies, Three dimensional bright spatial soliton collision and fusion in a
saturable nonlinear medium. Phys. Rev. Lett. 76, 2698–2701 (1996).


\bibitem{GS1} K. Li, P. Kevrekidis, D. J. Frantzeskakis, C. E. R\"{u}ter. and D. Kip,   Revisiting the-symmetric trimer: bifurcations, ghost states and associated dynamics. J. Phys. A: Math. Theor. 46, 375304 (2013).

\bibitem{GS2} H. Cartarius, D. Haag, D. Dast. and G. Wunner,  Nonlinear Schr\"odinger equation for a-symmetric delta-function double well. J. Phys. A Math. Theor., 45, 444008 (2012).

\bibitem{GS3} H. Susanto, R. Kusdiantara, N. Li, O. B. Kirikchi, D. Adzkiya, E. R. M. Putri. and T. Asfihani,  Snakes and ghosts in a parity-time-symmetric chain of dimers. Phys. Rev. E 97, 062204 (2018).

\bibitem{Yang} J. Yang,  Nonlinear Waves in Integrable and Nonintegrable Systems (SIAM, 2010).


\end{thebibliography}
\end{document}